\documentclass[final,3p,times]{elsarticle}

\usepackage{amssymb}
\usepackage{amsmath}
\usepackage[most]{tcolorbox}

\usepackage{ulem}
\usepackage{color}

\journal{Nuclear Physics B}

\newcommand{\Fourier}{Fourier }

\newcommand{\Heisenberg}{Heisenberg }

\let\oriDiv\div
\usepackage{physics}
\usepackage{physics2}
\AtBeginDocument{
  \let\div\oriDiv
}
\usephysicsmodule{ab}
\usephysicsmodule{diagmat}
\usephysicsmodule{braket}
\usepackage{etoolbox}
\AtBeginDocument{\undef{\qty}}
\usepackage{amsmath}
\usepackage{amssymb}
\usepackage{amsfonts}
\usepackage{latexsym}
\usepackage{mathtools}
\usepackage{bm}
\usepackage{ifthen}
\usepackage[thicklines]{cancel}
\usepackage{ascmac}
\usepackage{accents}
\usepackage{tensor}
\usepackage{slashed}

\newcommand{\proptosim}{\mathrel{\vcenter{\offinterlineskip
  \hbox{\scalebox{0.95}{$\propto$}}\hbox{\kern+0.0ex\raisebox{-0.8ex}{\hspace{0.00em}\scalebox{1.00}{$\sim$}}}}}}

\renewcommand{\eqref}[1]{\text{Eq. (\ref{#1})}}

\title{Estimation of potential radius based on momentum distribution of a constituent particle} 

\tnotetext[t1]{RIKEN-iTHEMS-Report-26}

\begin{document}
\begin{frontmatter}

\author[ISCT]{Eisuke Kawamura} 
\ead{kawamura.e.ab@m.titech.ac.jp}

\author[ISCT,iTHEMS]{Kotaro Murakami} 
\author[ISCT]{Daisuke Jido} 

\affiliation[ISCT]{
  organization={Department of Physics, Institute of Science Tokyo},
addressline={Ookayama 2-12-1}, 
city={Meguro-ku},
postcode={152-8550}, 
state={Tokyo},
country={Japan}}

\affiliation[iTHEMS]{
    organization={RIKEN Center for Interdisciplinary Theoretical and
Mathematical Sciences (iTHEMS)},
    addressline={RIKEN},
    city={Wako},
    postcode={351-0198},
    country={Japan}}

\begin{abstract}
  We propose using the potential radius as a probe of the structure of hadrons, particularly to classify exotic hadrons as hadronic or quark composite states.
  In this study, we focus on the radius of the effective potential felt by each constituent particle.
  Using a simple model with a square-well potential, we demonstrate that the potential radius can be estimated from the momentum distribution of a constituent particle not only for deeply bound states but also for shallowly bound states.
  We find that the momentum-based quantity provides a more robust estimate of the potential radius in the shallow-binding regime.
  This is because the momentum-based length scale decreases to zero as the potential radius vanishes, whereas the RMS radius approaches a finite value set by the binding energy.
  As a result, the momentum distribution avoids the finite-intercept problem that can make the inverse estimate of the potential radius ill-defined.
  With future experimental data on the momentum distribution of the constituent nucleon in $\overline{K}NNN$ production at J-PARC, the potential radius may be determined within the present framework. 
\end{abstract}


\end{frontmatter}



\section{Introduction}\label{sec1}
Recent technological developments have made it possible to investigate the structure of hadrons quantitatively, particularly that of exotic hadrons.
For example, there are plans to systematically produce anti-kaonic nuclei at J-PARC.
Accordingly, theoretical discussions of the structure of exotic hadrons are attracting increasing attention.

In this work, we propose a classification scheme for the structure of hadrons, with exotic hadrons as the primary targets.
Exotic hadrons may be classified into two classes: hadronic and quark composite states.
In the context of compositeness, this classification is associated with the distinction between composite and elementary.
However, there is no direct correspondence between the two approaches.

To classify a hadron as either a hadronic composite state or a quark composite state, we suggest examining the radius of the effective potential in the system.
In this study, we focus on one constituent particle and classify the composite state into one of the two types in terms of the effective potential acting on that particle.
If the state is a hadronic composite, the interaction between the constituents may be described as an interaction via hadron exchange.
Such interactions typically act over distances on the nuclear scale.
One representative example is nuclei, which are typical hadronic systems and are formed by the nuclear force.
On the other hand, if the state is a quark composite, the interaction may be described in terms of quark degrees of freedom.
Such an interaction may also take place within the quark-confinement range.
This range is comparable to the size of a single hadron.
The effective interaction is not unique and depends on how the system is modeled.
Nevertheless, the resulting potential radius is expected to be stable at the order-of-magnitude level against details of the modeling.

As mentioned above, the quantity we focus on is the potential radius, not the spatial extent of the wave function.
We expect that the spatial extent of the wave function reflects a different aspect of the state from that reflected by the potential radius.
The spatial extent of the wave function is known to correlate strongly with the binding energy.
In particular, in the shallow-binding limit the wave function is extended independently of the potential.
Consequently, the wave function can extend far beyond the radius of the effective potential.

In this study, we show that $\sqrt{\aab{p^{2}}}$ is a useful quantity for estimating the potential radius.
To do so, we compare the root-mean-square (RMS) radius $\sqrt{\aab{r^{2}}}$ and the RMS momentum $\sqrt{\aab{p^{2}}}$ of a constituent particle.
We study their dependence on the potential radius in a simple quantum-mechanical model with a square-well potential.
In the shallow-binding regime, the spatial extent of the wave function can be much larger than the potential radius and is mainly governed by the binding energy.
Therefore, $\sqrt{\aab{r^2}}$ is not suitable for estimating the potential radius in this regime.
This difficulty can be avoided by using the momentum distribution.
As shown in the next section, the momentum-based length scale decreases to zero as the potential radius decreases to zero, whereas $\sqrt{\aab{r^{2}}}$ approaches a finite universal value in the shallow-binding limit.
Therefore, the momentum distribution provides a more suitable quantity for inferring the potential radius in this regime.

One of the possible applications of our classification is the $\overline{K}NNN$ state. 
Anti-kaonic nuclei are exotic hadrons whose internal structure is under debate.
Due to the large binding energy, the spatial extent of the wave function is expected to be much smaller than that of ordinary nuclei. 
This implies that anti-kaonic nuclei may have a structure distinct from that of ordinary nuclei.
A production experiment for the $\overline{K}NNN$ nucleus is planned at J-PARC~\citep{E80_PoS}. 
In that experiment, the momentum distribution of the constituent nucleon may be inferred from the two-nucleon absorption process~\citep{Kienle05}.
In this study, we demonstrate how the $\overline{K}NNN$ state can be classified by deriving square-well potential radii from assumed values of $\sqrt{\aab{p^{2}}}$.
With future experimental data on the momentum distribution, the potential radius may be determined.
This would allow us to classify the $\overline{K}NNN$ state within our classification scheme.


\section{Relation between potential radius and momentum distribution}
We consider a bound state in a three-dimensional square-well potential.
This model allows us to examine how the potential radius is reflected in the coordinate and momentum scales of a constituent particle.
We calculate the RMS radius $\sqrt{\aab{r^2}}$ and the RMS momentum $\sqrt{\aab{p^2}}$ from the ground-state solution of the Schrödinger equation and study their dependence on the potential radius $R$. 
We regard the mass $\mu$, the potential radius $R$, and the binding energy $b$ as input parameters, whereas the potential depth $V_0$ is determined by $R$ and $b$. 
To compare $\sqrt{\aab{p^{2}}}$ with $\sqrt{\aab{r^{2}}}$ on the same length scale, we introduce $r_p$\footnote{
  The quantity $r_{p}$ is defined such that 
  it is equal to $\sqrt{\aab{r^{2}}}$ if and only if the \Heisenberg uncertainty product $\sqrt{\aab{r^{2}}}\sqrt{\aab{p^{2}}}/\hbar$ takes the minimum value $3/2$,
} as
\begin{equation}
  r_p \equiv \frac{3}{2}\frac{\hbar}{\sqrt{\aab{p^2}}}.
\end{equation}
Introducing the binding-energy-dependent inverse length scale
$\varrho = \sqrt{2\mu b}/\hbar$, the quantities $\varrho\sqrt{\aab{r^2}}$, $\varrho r_p$, and $\varrho R$ are dimensionless.
The parameter $\varrho R$ characterizes how deeply the state is bound:
$\varrho R > 1$ corresponds to a deeply bound state, whereas $\varrho R < 1$ corresponds to a shallowly bound state.

\begin{figure}[htbp]
  \centering
  \includegraphics[width=0.325\textwidth]{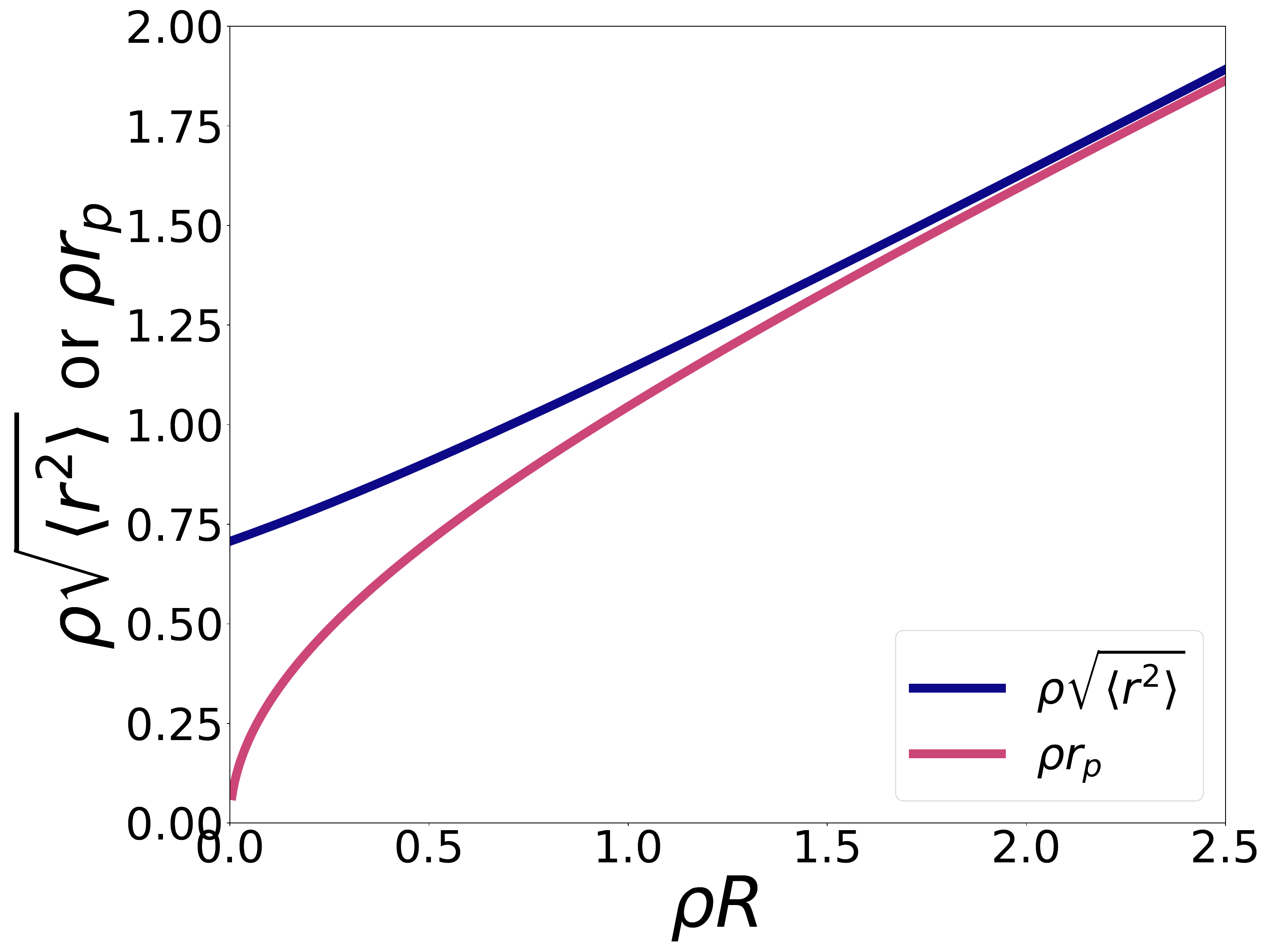}
  \includegraphics[width=0.325\textwidth]{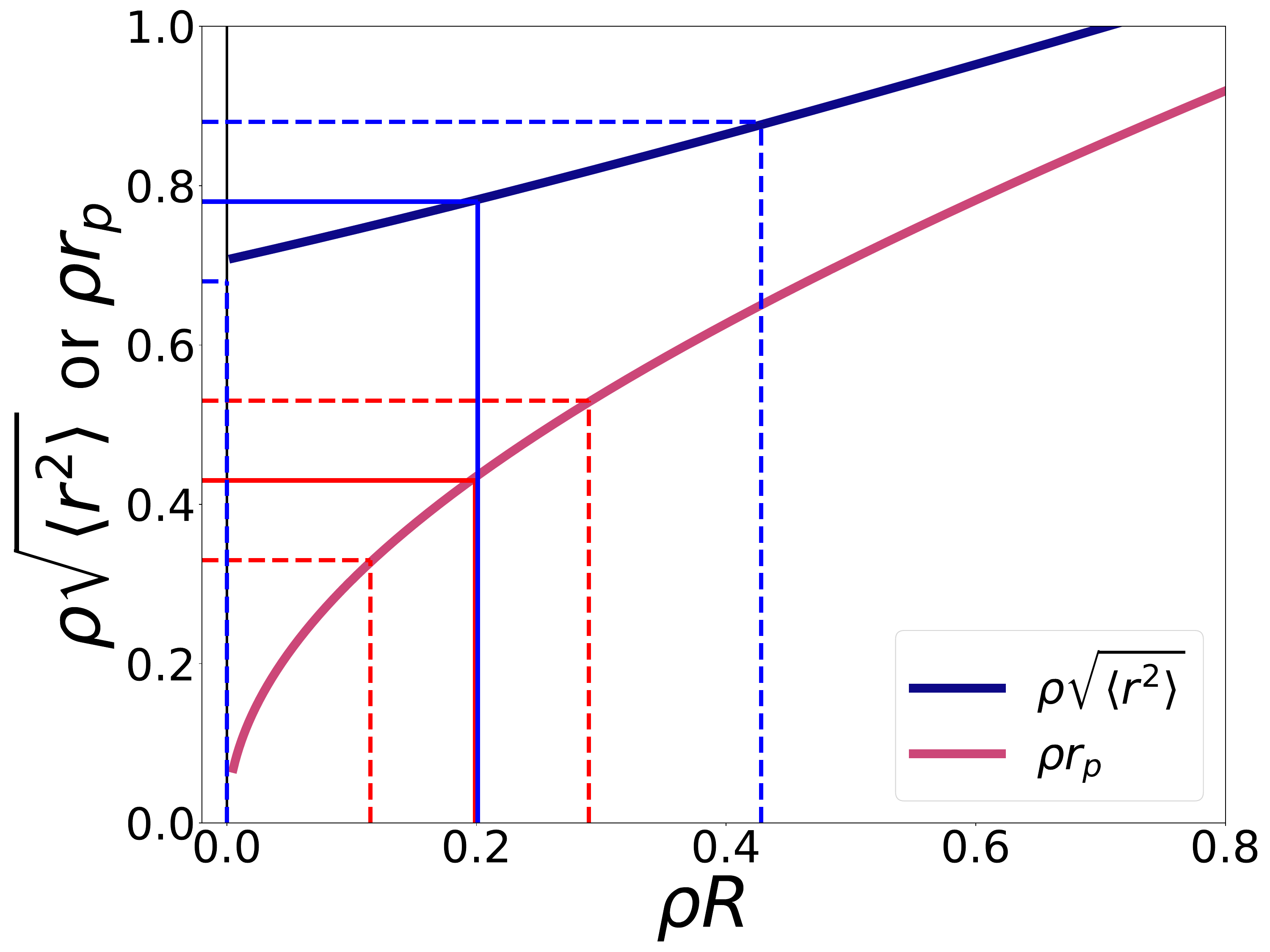}
  \includegraphics[width=0.325\textwidth]{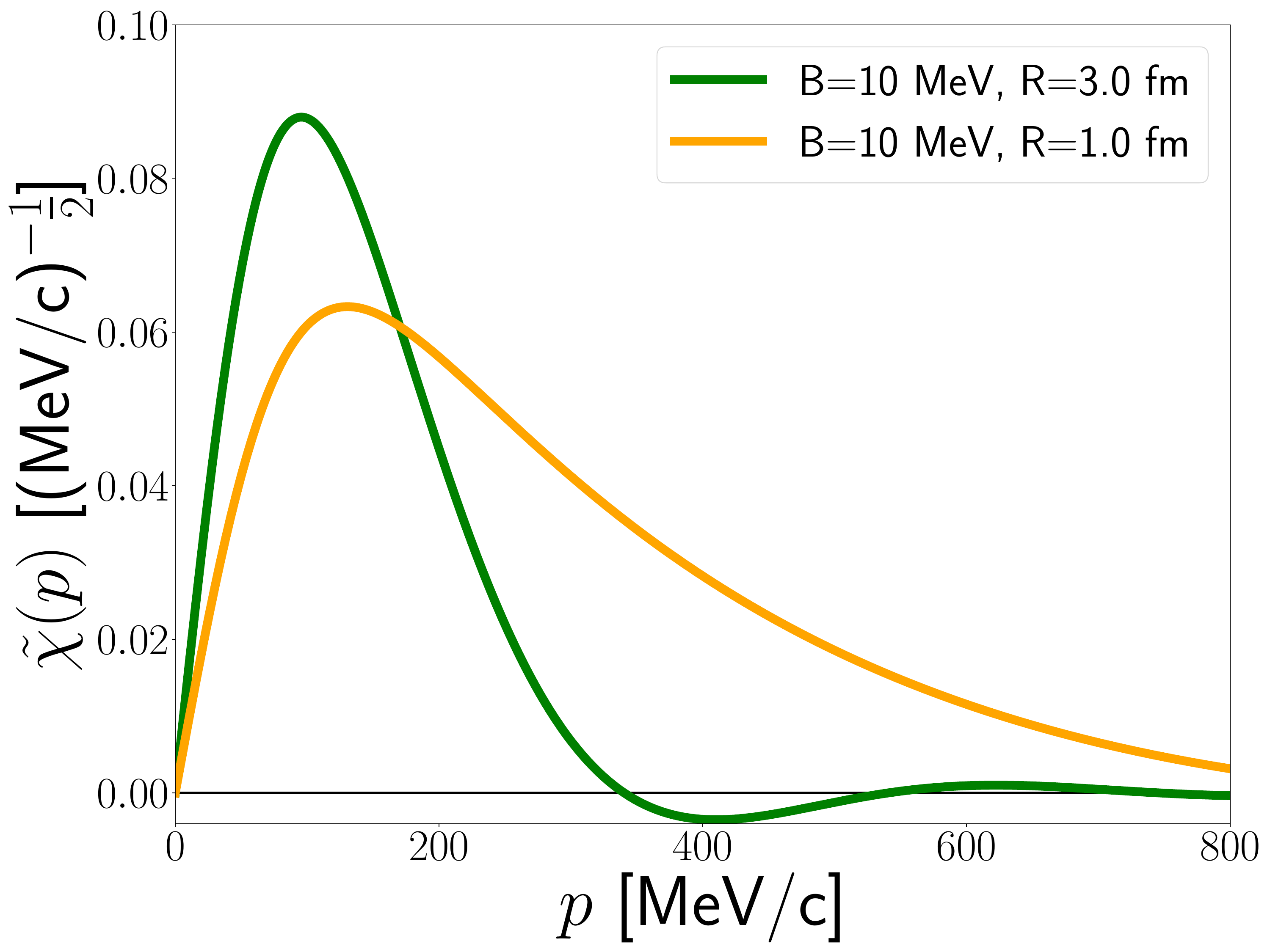}

  \caption{
    Left: 
    Dependence of $\varrho \sqrt{\aab{r^{2}}}$ and $\varrho r_{p}$ on $\varrho R$ for the bound state.
    Center: Enlarged plot of the left panel. 
    Horizontal dashed bands indicate symmetric error bands of 0.78 ± 0.10 for $\varrho\sqrt{\aab{r^{2}}}$ (blue) and 0.43 ± 0.10 for $\varrho r_{p}$ (red).
    Vertical dashed bands show the corresponding propagation of these errors to $\varrho R$.
    Right: 
    The \Fourier transform of the wave functions with the same binding energy but different potential radius.
  }
  \label{図:rhoR-rhor}
\end{figure}

The left panel of Fig.~\ref{図:rhoR-rhor} shows the $\varrho R$ dependence of $\varrho\sqrt{\aab{r^2}}$ and $\varrho r_p$. 
In the deeply bound regime, $\varrho R > 1$, both quantities show similar behavior.
In contrast, in the shallow-binding regime, $\varrho R < 1$, their behaviors are qualitatively different: $\varrho\sqrt{\aab{r^2}}$ approaches a finite value, whereas $\varrho r_p$ decreases toward zero.
This difference is important because it determines whether the inverse estimate of $\varrho R$ is well defined near the shallow-binding limit.

For deeply bound states, as seen in the region $\varrho R > 1$, in the left panel of Fig.~\ref{図:rhoR-rhor}, $\varrho\sqrt{\aab{r^2}}$ and $\varrho r_p$ show similar behavior.
At the same time, both quantities linearly depend on $\varrho R$.
This can be seen from the expansion in powers of $1/(\varrho R)$:
\begin{subequations}
  \begin{align}
    \aab{r^{2}} &= \pab{\frac{1}{3}-\frac{1}{2\pi^{2}}}R^{2}
    \pab{
    1 + \frac{2}{\varrho R} + \order{\frac{1}{\pab{\varrho R}^{2}}}
    },
    \\
    r_{p}^{2} &= 
    \frac{9}{4}\frac{R^{2}}{\pi^{2}}\pab{1 + \frac{3}{\varrho R} + \order{\frac{1}{\pab{\varrho R}^{2}}}}.
  \end{align}
\end{subequations}
The leading-order terms of both $\aab{r^2}$ and $r_{p}^{2}$ are proportional to $R^2$.

These behaviors have the following simple physical interpretation.
In the deeply bound regime, the wave function is localized within the potential region.
Therefore, the coordinate-space scale $\sqrt{\aab{r^2}}$ is directly controlled by the potential radius $R$.
In addition, we find that the uncertainty product $\sqrt{\aab{r^2}}\sqrt{\aab{p^2}}/\hbar$ approaches a value close to $3/2$, the lower bound of the Heisenberg uncertainty principle.
Consequently, the length scale $r_p$, defined from the RMS momentum, becomes comparable to $\sqrt{\aab{r^2}}$.
Thus, in the deeply bound regime, both $\sqrt{\aab{r^2}}$ and $r_{p}$ can be used to estimate the potential radius.

In the shallow-binding regime ($\varrho R < 1$), $\sqrt{\aab{r^{2}}}$ loses the correspondence to $R$.
As shown in the left panel of Fig.~\ref{図:rhoR-rhor}, 
with decreasing $\varrho R$,
$\varrho \sqrt{\aab{r^{2}}}$ approaches an $R$-independent value.
This behavior reflects low-energy universality.
In this regime, the wave function extends far outside the potential region, and its spatial extent is mainly governed by the binding energy rather than by the potential radius.
Consequently, $\varrho\sqrt{\aab{r^2}}$ no longer reflects the information on $\varrho R$.

On the other hand, $r_p$ retains an explicit dependence on $R$ in the shallow-binding regime.
As seen in the left panel of Fig.~\ref{図:rhoR-rhor}, $\varrho r_p$ decreases to zero as $\varrho R$ approaches zero.
Thus, unlike $\varrho\sqrt{\aab{r^2}}$, the momentum-based length scale has no finite intercept at the edge of the model domain.
This behavior is seen from the expansion in powers of $\varrho R$:
\begin{subequations}\label{式:shallow limit}
  \begin{align}
    \aab{r^{2}}
    &= \frac{1}{2\varrho^{2}}\bab{1 + \varrho R + \order{\pab{\varrho R}^{2}}},\label{式:shallow limit1} 
    \\
    r_{p}^{2}
    &= \frac{9}{\pi^{2}}\frac{R}{\varrho}
    \pab{1 + \pab{1- \frac{8}{\pi^{2}}}\varrho^{}R^{} + \order{\pab{\varrho R}^{2}}},\label{式:shallow limit2}
  \end{align}
\end{subequations}
At leading order, $\sqrt{\aab{r^2}}$ does not depend on $R$.
In contrast, $r_{p}^{2}$ retains a dependence on $R$ even at leading order.

This difference becomes important when the relation is inverted to estimate the potential radius from an experimentally extracted quantity.
To illustrate this point quantitatively, suppose first that $\varrho \sqrt{\aab{r^{2}}}$ is obtained as 0.78.
According to the relation shown in the center panel of Fig.~\ref{図:rhoR-rhor}, this value corresponds to $\varrho R = 0.2$.
If we assign a symmetric error of $\pm 0.10$ to this value, namely $\varrho\sqrt{\aab{r^2}} = 0.78 \pm 0.10$, the corresponding range of $\varrho R$ becomes approximately $0 < \varrho R < 0.4$.
We then consider the case where $\varrho r_p$ is obtained as $0.43$, which also corresponds to $\varrho R = 0.2$.
Assigning the same symmetric error, $\varrho r_p = 0.43 \pm 0.10$, gives approximately $0.1 < \varrho R < 0.3$.

From the demonstration of error propagation, the essential advantage of $\varrho r_p$ is that its relation to $\varrho R$ starts from the origin.
By contrast, $\varrho\sqrt{\aab{r^2}}$ has a finite $y$-intercept in the shallow-binding limit.
If an experimentally extracted value, or its error band, lies below or overlaps this intercept, the inversion to $\varrho R$ becomes ill-defined within the present model, and the lower bound of $\varrho R$ collapses to zero.
In contrast, $\varrho r_p$ reaches zero only at $\varrho R=0$.
Therefore, any positive value of $\varrho r_p$ can be associated with a positive potential radius within the model.
Furthermore, even for the same size of the error in the measured quantity, $\varrho r_p$ constrains $\varrho R$ more tightly than $\varrho\sqrt{\aab{r^2}}$.
Therefore, $\varrho r_{p}$ is a better indicator for estimating the potential radius at the order-of-magnitude level.

The contrast between $\sqrt{\aab{r^{2}}}$ and $r_{p}$ can be understood by considering how the wave function behaves in coordinate and momentum space.
In the shallow-binding limit, the wave function becomes broader regardless of the potential shape.
As a result, its spatial extent becomes large almost independently of the potential radius.
On the other hand, the high-momentum tail of the wave function is controlled by $R$.
As seen in the right panel of Fig.~\ref{図:rhoR-rhor}, the wave function with smaller $R$ exhibits a more pronounced high-momentum tail. 
Therefore, larger RMS momentum (or smaller $r_{p}$) corresponds to a shorter potential radius.
This is the physical reason why the momentum distribution contains information on the potential radius even when the RMS radius is dominated by the binding energy.

The simple correspondence between $\sqrt{\aab{r^2}}$ and $r_p$ does not hold in the shallow-binding regime.
In the deep-binding regime, $\sqrt{\aab{r^2}}$ and $r_p$ are similar because the product approaches $\sqrt{\aab{r^2}}\sqrt{\aab{p^2}}/\hbar \simeq 3/2$.
This approximate relation is often assumed to hold too generally.
However, we find that this approximation does not hold in the shallow-binding regime.
The product deviates from $3/2$ and grows as $\sqrt{\aab{r^2}}\sqrt{\aab{p^2}}/\hbar \sim \order{\frac{1}{\varrho R}}$.

\section{Applications}

In this section, we demonstrate how the potential radius can be obtained from the momentum distribution of a constituent particle.
As shown in the previous section, the length scale $r_p$, defined from the RMS momentum, provides a well-defined way to infer the potential radius even in the shallow-binding regime.
Therefore, in the following examples, we estimate $R$ using $r_p$.
We consider two systems: the deuteron and the $KNNN$ nucleus.
For each system, we specify the mass $\mu$, the binding energy $b$, and $r_p$ defined from $\sqrt{\aab{p^2}}$, and then determine $R$ using the analytical solution of the square-well potential.

We first demonstrate the applicability of this model to a hadronic composite state using deuteron data for a well-known hadronic system.
We have well-organized deuteron models, which provide the wave function of the relative motion of the nucleons. 
With this wave function, we can estimate the momentum distribution.
Adopting the momentum distribution from the CD-Bonn potential~\citep{R.Machleidt_2001} yields $\sqrt{\aab{p^{2}}}$ of $121\ \mathrm{MeV}/ c$ ($r_{p} = 2.4\ \mathrm{fm}$).
We apply this value to the square-well potential.
The reduced mass is $m_N/2 = 469\ \mathrm{MeV}/c^2$, and the deuteron binding energy is $B_d = 2.2\ \mathrm{MeV}$ ($\varrho = 0.23\ \mathrm{fm^{-1}}$).
We then obtain $\varrho R = 0.33$, that is, $R = 1.44\ \mathrm{fm}$.
This value is larger than the size of a single hadron and lies on the nuclear length scale. 
It is also of the same order as the range of the long-range potential mediated by pion exchange. 
Therefore, this length scale supports a hadronic composite, rather than quark-composite, interpretation.


We next apply the method to the $\overline{K}NNN$ nucleus.
The structure of the $\overline{K}NNN$ nucleus can be probed through the momentum distribution of a constituent nucleon.
This momentum distribution may be inferred from one of the decay processes, $\overline{K}NNN \rightarrow \Lambda + N + N$.
In the decay process, we expect that two-nucleon absorption of $\overline{K}$ is dominant and one of the nucleons acts as a spectator~\citep{Kienle05}.
Thus, the momentum distribution of the spectator nucleon can be regarded as the relative momentum distribution of the spectator nucleon and the rest of the constituents in the bound state.
With this momentum distribution, we evaluate the potential radius of an effective two-body system composed of $N$ and $\overline{K}NN$.
For this effective two-body system, the reduced mass is obtained from the masses of $N$ and $\overline{K}NN$ as $\mu = m_N M_{\overline{K}NN}/(m_N + M_{\overline{K}NN})$.
Using $m_N=940\,\mathrm{MeV}/c^2$ and $M_{\overline{K}NN}=2330\,\mathrm{MeV}/c^2$, we take the reduced mass to be $670\,\mathrm{MeV}/c^2$.
We also adopt $b = 20\,\mathrm{MeV}$ ($\varrho = 0.83\ \mathrm{fm^{-1}}$) as an illustrative binding-energy parameter for the effective $\overline{K}NN$-$N$ relative motion.
This value $b = 20 \, \mathrm{MeV}$ is chosen as a representative scale between $B_{\overline{K}N}=15\,\mathrm{MeV}$ and $B_{\overline{K}NN}/2=21\,\mathrm{MeV}$.

For these parameters, we examine how the potential radius changes for different values of $\sqrt{\aab{p^{2}}}$.
We compare the cases $\sqrt{\aab{p^{2}}} = 300\ \mathrm{MeV}/c$ ($r_{p} = 1\ \mathrm{fm}$) and $\sqrt{\aab{p^{2}}} = 150 \ \mathrm{MeV}/c$ ($r_{p} = 2\ \mathrm{fm}$).
For $\sqrt{\aab{p^2}}=300\,\mathrm{MeV}/c$, we obtain $\varrho R=0.65$, that is $R=0.78\,\mathrm{fm}$.
Since this radius is comparable to the size of a nucleon at the order-of-magnitude level, the result clearly points to a quark-composite interpretation.
For $\sqrt{\aab{p^2}}=150\,\mathrm{MeV}/c$, we obtain $\varrho R=2.0$, that is $R=2.5\,\mathrm{fm}$.
This radius is well larger than the size of a nucleon at the order-of-magnitude level, the result clearly points to a hadronic-composite interpretation.
These examples demonstrate the usefulness of measuring $\sqrt{\aab{p^{2}}}$.
Such a measurement can provide information on whether the state is better described as a quark composite or a hadronic composite.

\section{Summary}
We propose examining the potential radius of the system as a probe of the structure of exotic hadrons.
In this study, we focus on the radius of the effective potential felt by each constituent particle.
We show that $\sqrt{\aab{p^{2}}}$ of a constituent particle is a key indicator for estimating the potential radius.
Unlike the spatial extent of the wave function, this method is effective even for shallowly bound states.
With the future J-PARC experiment aimed at producing the $\overline{K}NNN$ nucleus, the potential radius of the $\overline{K}NNN$ nucleus may be accessed using our method.
This method is model-dependent because it relies on the definition of the potential radius and on the assumed potential shape.
However, our calculations suggest that the order of magnitude of the potential radius does not differ significantly among several models.
A more detailed evaluation of this model dependence is left for future work.

\section{Acknowledgements}
This work was partially supported 
by the International Physics Leadership Program at the Department of Physics, Institute of Science Tokyo 
and by the Grants-in-Aid for Scientific Research from JSPS (JP22H04917, JP23K03427 and JP25K07315).



\bibliographystyle{elsarticle-harv} 
\bibliography{references}     
\vspace{1em}
[4] E. Kawamura, K. Murakami, D. Jido, in preparation







\end{document}